%% file: RtaumuLetter.tex
\documentclass[twocolumn,showpacs,aps,prl,superscriptaddress,floatfix]{revtex4}

\usepackage{graphicx}
\usepackage{verbatim}
\usepackage{dcolumn}
\usepackage{amsmath}
\usepackage{epsfig}
\usepackage{siunitx}

\newcommand{\BaBarYear}      {20}
\newcommand{\BaBarNumber}    {002}
\newcommand{\BaBarType}      {PUB}  
\newcommand{\SLACPubNumber}  {17527}

\input babarsym

\def\Br{{\cal B}}
\def\Rtm{\ensuremath{{\cal R}_{\tau\mu}^{\Upsilon(3S)}}\xspace}
\DeclareSIUnit\lspeed{\text{$c$}}
\def\kkmc {\mbox{\tt KKMC}\xspace}
\def\bhwide {\mbox{\tt BHWIDE}\xspace}
\def\photos {\mbox{\tt PHOTOS}\xspace}
\def\geant4 {\mbox{\tt GEANT4}\xspace}

\begin{document}
\pagestyle{plain}

\begin{flushleft}
\babar-\BaBarType-\BaBarYear/\BaBarNumber \\
SLAC-PUB-\SLACPubNumber\\
\end{flushleft}

\begin{flushleft}
\end{flushleft}

\title{\large \bf Precision measurement of the $\Br(\Upsilon(3S)\to\tau^+\tau^-)/\Br(\Upsilon(3S)\to\mu^+\mu^-)$ ratio}
\pacs{14.60.-z, 14.60.Fg, 14.60.Ef, 14.40.Nd, 14.40.-n, 13.20.-v, 11.30.Hv, 11.30.-j}
\begin{abstract}
  We report on a precision measurement of the ratio $\Rtm =
  \Br(\Upsilon(3S)\to\tau^+\tau^-)/\Br(\Upsilon(3S)\to\mu^+\mu^-)$
  using data collected with the \babar\ detector at the SLAC PEP-II
  $e^+e^-$ collider. The measurement is based on a
  \SI{28}{\per\femto\barn} data sample collected at a center-of-mass
  energy of \SI{10.355}{\giga\electronvolt} corresponding to a sample
  of 122 million $\Upsilon(3S)$ mesons.  The ratio is measured to be
  $\Rtm = 0.966 \pm 0.008_\text{stat} \pm 0.014_\text{syst}$ and is in
  agreement with the Standard Model prediction of 0.9948 within 2
  standard deviations. The uncertainty in \Rtm is almost an order of
  magnitude smaller than the only previous measurement.
\end{abstract}
\input authors_feb2020_frozen.tex
\maketitle
In the Standard Model (SM) the width of a spin~1 bound state of a
quark and antiquark decaying into a charged lepton-antilepton pair in
the absence of radiation effects has been known since the model's
inception~\cite{VanRoyen:1967nq} to be:
$$
\Gamma_{\ell\ell} = 4\alpha^2 e_q^2
\frac{|\Psi(0)|^2}{M^2}\left(1+2\frac{m_{\ell}^2}{M^2}\right)
\sqrt{1-4\frac{m_{\ell}^2}{M^2}},
$$ where $\Gamma_{\ell\ell}$ is the decay width to two leptons of
flavor $\ell$ (e.g., muon or $\tau$ lepton), $\alpha$ is the fine
structure constant, $e_q$ is the quark charge, $\Psi(0)$ is the value
of the radial wave function evaluated at the origin, $M$ is the
resonance mass, and $m_\ell$ is the lepton mass. The ratio of widths
to final-state leptons with different flavor is free of hadronic
uncertainties, and for heavy spin~1 resonances, such as the family of
the $b\bar{b}$ bound states $\Upsilon(nS)$ mesons, differs from unity
only by a small mass correction. Consequently, leptonic decays of the
$\Upsilon(nS)$ mesons are good candidates to test SM predictions and
to search for phenomena beyond the SM.  For example, the CP-odd Higgs
boson $A^0$ proposed in Ref.~\cite{SanchisLozano:2003ha} couples more
strongly to heavier fermions.  This would introduce the $\Upsilon(nS)
\to \gamma A^0 \to \gamma\tau^+\tau^-$ decay chain with a rate
substantially higher than that of the $\Upsilon(nS) \to \gamma A^0\to
\gamma\mu^+\mu^-$ chain and result in a larger value of the ratio
$\Rtm =
\Br(\Upsilon(3S)\to\tau^+\tau^-)/\Br(\Upsilon(3S)\to\mu^+\mu^-)$ than
predicted in the SM. The only measurement to date of that ratio was
made by the CLEO collaboration, $\Rtm = 1.05 \pm 0.08 \pm
0.05$~\cite{Besson:2006gj}.  It has also been
remarked~\cite{Aloni:2017eny} that measuring this ratio could shed
light on the suggestion for new physics seen in $\Gamma(B\rightarrow
D^{(*)} \tau \nu)/\Gamma(B\rightarrow D^{(*)} (e/\mu)
\nu)$~\cite{Amhis:2016xyh}.  A new precise measurement will further
constrain new physics models.

In this Letter we present a precision measurement of the ratio \Rtm
using a novel technique to discriminate between resonant and
non-resonant (i.e., continuum) dimuon production based on differences
in the dimuon mass distributions associated with initial state
radiation (ISR).  In the resonant process, $e^+e^- \to \Upsilon(3S)\to
\mu^+\mu^-$, ISR is heavily suppressed compared to the non-resonant,
$e^+e^- \to \mu^+\mu^-$, process.  Details of how we estimate the
non-$\Upsilon(3S)$ contribution to the dimuon sample using this
technique are described below.  We account for the number of
non-resonant $\tau^+\tau^-$ events using information from the
continuum values of the number of dimuons together with the ratio of
the selected number of $\tau^+\tau^-$ to $\mu^+\mu^-$ events in the
$\Upsilon(4S)$ data control sample, corrected for center-of-mass
dependent phase-space effects.  This method ensures that the measured
ratio is fully inclusive of radiation effects and does not require a
precise luminosity determination.

The data samples used for these studies were collected with the
\babar\ detector at the \pep2 asymmetric-energy $e^+e^-$ collider at
the SLAC National Accelerator Laboratory. The \babar\ experiment
collected data between 1999 and 2008 at center-of-mass energies of the
$\Upsilon(4S)$, $\Upsilon(3S)$, and $\Upsilon(2S)$ resonances, as well
as at nonresonant energies. The PEP-II positron beam energy was
\SI{3.1}{\GeV}, while the electron beam energy was \SI{8.6}{\GeV} at
the $\Upsilon(3S)$ and \SI{9.0}{\GeV} at the $\Upsilon(4S)$, resulting
in different boosts of the final-state system and different detector
acceptances in the center-of-mass frame. We measure the ratio
$R_{\tau\mu}^{\Upsilon(3S)}$ using a sample of 122 million
$\Upsilon(3S)$ decays corresponding to an integrated luminosity of
\SI{27.96}{\per\femto\barn}~\cite{lumi} collected at $\sqrt{s}=\SI{10.355}{\GeV}$
during 2008 (referred to as Run~7), where $\sqrt{s}$ is the
center-of-mass energy.  We also employ three data control samples:
data collected at the $\Upsilon(4S)$ in 2007 (referred to as Run~6)
corresponding to \SI{78.3}{\per\femto\barn}, data taken \SI{40}{\MeV}
below the $\Upsilon(4S)$ resonance (termed ``off-resonance'')
corresponding to \SI{7.75}{\per\femto\barn}, and data taken
\SI{30}{\MeV} below the $\Upsilon(3S)$ resonance corresponding to
\SI{2.62}{\per\femto\barn}.  All data used in this analysis were
collected with the same detector configuration after the last major
upgrade in 2007.  These data control samples are used to evaluate
properties of the background, to study systematic effects, and to
calculate corrections to Monte Carlo (MC) based efficiencies.  We
employ a blind analysis~\cite{ref:blindanalysis} in which only a small
subset of \SI{2.41}{\per\femto\barn} of the total $\Upsilon(3S)$
sample is used in the pre-unblinding stage during which selection
criteria are optimized.

\babar\ was a general purpose detector and is described in detail
elsewhere~\cite{ref:aubert,ref:NIMUpdate}. Its magnetic spectrometer,
used to measure momenta of charged particles, comprised a 5-layer
silicon vertex tracker (SVT) surrounded by a 40-layer cylindrical
drift chamber (DCH) placed inside a \SI{1.5}{\tesla} superconducting
solenoid with its axis aligned nearly parallel to the $e^+e^-$ beams.
Charged hadron identification was performed by using ionization
measurements in the SVT and DCH and by using a ring-imaging Cherenkov
detector (DIRC), which formed a cylinder surrounding the DCH.  The
\babar\ electromagnetic calorimeter (EMC), composed of an array of
6580 CsI(Tl) crystals located between the DIRC and the solenoid, was
used to measure energies and directions of photons as well as to
identify electrons. Muons and neutral hadrons were identified by
arrays of resistive plate chambers or limited steamer-tube detectors
inserted into gaps in the steel of the Instrumented Flux Return (IFR)
of the magnet. An upgrade of the IFR was completed in 2007 prior to
Run~6.

The major irreducible background process is continuum dilepton
production.  The \kkmc event generator~\cite{ref:ward} is used to
simulate continuum $\mu^+\mu^-$ and $\tau^+\tau^-$ production taking
into account radiative effects. For the Bhabha process the
\bhwide~\cite{ref:BHWIDE} event generator is employed. The \evtgen
generator~\cite{ref:lange} is used to simulate hadronic continuum
events and generic $\Upsilon(3S)$ decays, with the final-state
radiation effects modeled by means of the \photos
package~\cite{photos}. The simulated $\mu^+\mu^-$, $\tau^+\tau^-$, and
generic $\Upsilon(3S)$ samples correspond to roughly twice the number
of events in the $\Upsilon(3S)$ dataset, while the Bhabha sample
corresponds to roughly half the number of events.  In addition, the
$\Upsilon(3S)\to\mu^+\mu^-$ and $\Upsilon(3S)\to\tau^+\tau^-$ signal
decays are simulated using the \kkmc generator with ISR turned off.
Thus the same MC generator, \kkmc, is employed for both the signal and
continuum, which enables a consistent evaluation of the corrections to
the discrepancies between data and MC.  This signal MC sample is about
three times the size of the data sample.  Particle interactions with
the detector and its response are modeled within the \geant4
framework~\cite{ref:agostine}.

As mentioned above, the selection criteria are developed using a
\SI{2.41}{\per\femto\barn} $\Upsilon(3S)$ subsample (approximately
one-tenth of the total data) to avoid possible biases. The dimuon
candidate requires two and only two reconstructed high momentum
collinear ($<\SI{20}{\degree}$) charged particles in the center-of-mass
frame with opposite charges and with associated energy depositions in
the EMC consistent with the muon hypothesis.  We apply a polar angle
selection of
$\SI{37}{\degree}\lesssim\theta_{-}\lesssim\SI{143}{\degree}$ and
$\SI{33}{\degree}\lesssim\theta_{+}\lesssim\SI{147}{\degree}$, where
$\theta_{-}$ and $\theta_{+}$ are polar angles in the center-of-mass
frame of negative and positive muon candidates respectively.  This
selection provides the same efficiency at the borders of the sensitive
volume for different boost values at different energies in the
laboratory frame. Misidentified Bhabha events are additionally
suppressed with a requirement that at least one of the muon candidates
in an event has a response in the IFR. The scaled invariant mass
$M_{\mu\mu}/\sqrt{s}$ of the two muons must be in the range
$0.8<M_{\mu\mu}/\sqrt{s}<1.1$. These selection criteria provide a
dimuon sample with 99.9\% purity, according to MC studies.

We consider $\tau$-pairs where both taus have a single charged
particle in their decay, where one of the charged particles is an
electron and the other is not an electron.  The $\tau^+\tau^-$
candidate selection proceeds by requiring two and only two
reconstructed tracks with opposite charges in the event. One of the
tracks is required to be identified as an electron based on energy
deposition in the tracking system and EMC, whereas the other track
must fail the same electron selection requirements.  Backgrounds are
further suppressed by requiring the angle between the two tracks to be
greater than \SI{110}{\degree} in the center-of-mass frame. The total
energy registered in the EMC must be less than \SI{70}{\percent} of
the initial $e^+e^-$ energy in the laboratory frame.  The
acollinearity between the two tracks in the azimuthal plane must be
greater than \SI{3}{\degree}.  The missing mass, $M_\text{miss}$,
which is based on the two tracks and up to the ten most energetic
clusters in the EMC identified as photons, must satisfy the
requirement that $|M^2_\text{miss}/s|>0.01$.  The missing momentum
vector must point to the sensitive part of the detector, defined as
$|\cos{\theta_\text{miss}}|<0.85$ in the center-of-mass frame. To
further suppress the Bhabha background, the acollinearity angle
between the non-electron track and the combination of the identified
electron track and the most energetic photon must be greater than
\SI{2}{\degree} in both azimuthal and polar angles in the
center-of-mass frame.

A large fraction of the background comes from two-photon processes
where tracks have low transverse momenta.  Since this region is also
populated by the signal $\tau^+\tau^-$ events, a two-dimensional cut
on the transverse momentum of the positive lepton $vs$ that of the
negative lepton is developed to remove the background and maintain an
acceptable efficiency for the signal events.  These selection criteria
provide a $\tau^+\tau^-$ sample with 99\% purity, according to MC
studies.

The \SI{2.62}{\per\femto\barn} $\Upsilon(3S)$ off-resonance and
\SI{7.75}{\per\femto\barn} $\Upsilon(4S)$ off-resonance samples are
used to correct for differences between MC and data
$\tau^+\tau^-/\mu^+\mu^-$ selection efficiency ratios. For the
experimental data and their corresponding MC samples, the number of
dilepton candidates (MC scaled to the data luminosity) and
corresponding efficiency corrections are shown in
Table~\ref{table:nn}.  For the $\Upsilon(3S)$ and $\Upsilon(4S)$
off-resonance data samples, the $N_{\tau\tau}/N_{\mu\mu}$ dilepton
candidate ratios are $0.11665 \pm 0.00029$ and $0.11647 \pm 0.00017$,
respectively.  These are in excellent agreement and show that for
these selections the efficiency ratio does not depend on the
center-of-mass energy or the different boost associated with the two
samples.  The corresponding MC samples show the same behavior and
allow us to extract data-driven corrections to the MC efficiency
ratio. The average correction to the MC efficiency ratio between
samples is $C_\text{MC} =
(\varepsilon_{\tau\tau}/\varepsilon_{\mu\mu})^\text{data}/(\varepsilon_{\tau\tau}/\varepsilon_{\mu\mu})^\text{MC}
= 1.0146 \pm 0.0016$.

\begin{table}[htbp!]
  \caption{\label{table:nn} The numbers of dilepton candidates in
    \SI{2.62}{\per\femto\barn} $\Upsilon(3S)$ and
    \SI{7.75}{\per\femto\barn} $\Upsilon(4S)$ off-resonance data and
    MC samples and the correction for data and MC efficiency
    discrepancies. The numbers of MC events are scaled according to the
    measured luminosity.}  \centering
  \begin{tabular}{c|c|c|c|c|c}
    \hline\hline
    Sample & $N_{\mu\mu}^\text{data}$ & $N_{\mu\mu}^\text{MC}$ & $N_{\tau\tau}^\text{data}$ & $N_{\tau\tau}^\text{MC}$ & $\dfrac{N_{\tau\tau}^\text{data}/N_{\mu\mu}^\text{data}}{N_{\tau\tau}^\text{MC}/N_{\mu\mu}^\text{MC}}$\\
    \hline
    $\Upsilon(3S)$ & 1,538,569 & 1,554,208 & 179,466 & 178,569 & $1.015 \pm 0.003$\\
    $\Upsilon(4S)$ & 4,422,407 & 4,398,983 & 515,067 & 505,133 & $1.014 \pm 0.002$\\
    \hline\hline
  \end{tabular}
\end{table}

\begin{figure}[htbp!]
  \centering
  \includegraphics[width=0.5\textwidth]{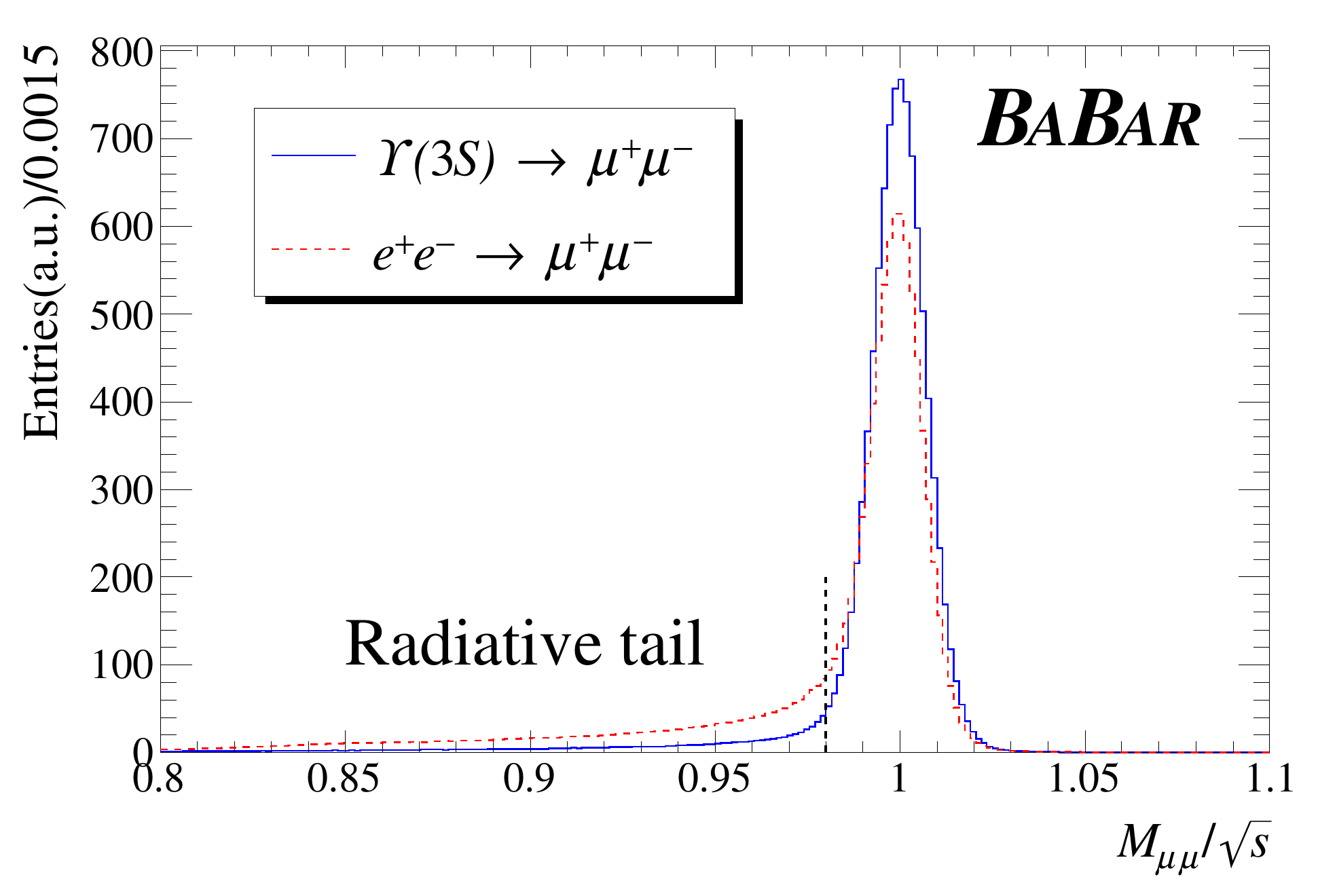}
  \caption{\label{fig:radtail} Comparison of $M_{\mu\mu}/\sqrt{s}$
    distributions for the continuum production $e^+e^-\to\mu^+\mu^-$
    in data at $\Upsilon(4S)$ off-resonance energy and $\Upsilon(3S)\to
    \mu^+\mu^-$ decays in MC, where only final-state radiation is
    expected. The distributions are normalized to the same number of
    events. The vertical dashed line shows the border $M_{\mu\mu}/\sqrt{s}=0.98$.}
\end{figure}

The method to discriminate between $\Upsilon(3S)\to\mu^+\mu^-$ decays
and the continuum production $e^+e^-\to\mu^+\mu^-$ is based on the
fact that the $\Upsilon(3S)$ resonance is very narrow and thus the ISR
effects are highly suppressed for the signal, but not the continuum
background.  If the ISR photons have an energy greater than a few MeV
(an amount associated with the PEP-II beam energy spread), then the
$e^+e^-$ interaction energy is too low to form the $b\bar{b}$ bound
state. This effect results in a significant difference in the
radiative tail of the $M_{\mu\mu}$ distribution for the continuum and
resonance production processes for reconstructed dimuon candidates, as
shown in Fig.~\ref{fig:radtail}.  About 23\% of the continuum
candidates are in the low mass radiative tail region
($M_{\mu\mu}/\sqrt{s}<0.98$; $3\sigma$ of invariant mass resolution
corresponds to approximately 0.02 in these units) whereas for the
resonance decays this number is only 7\%, and is associated with final
state radiation effects.

\begin{figure}[htbp!]
  \includegraphics[width=0.5\textwidth]{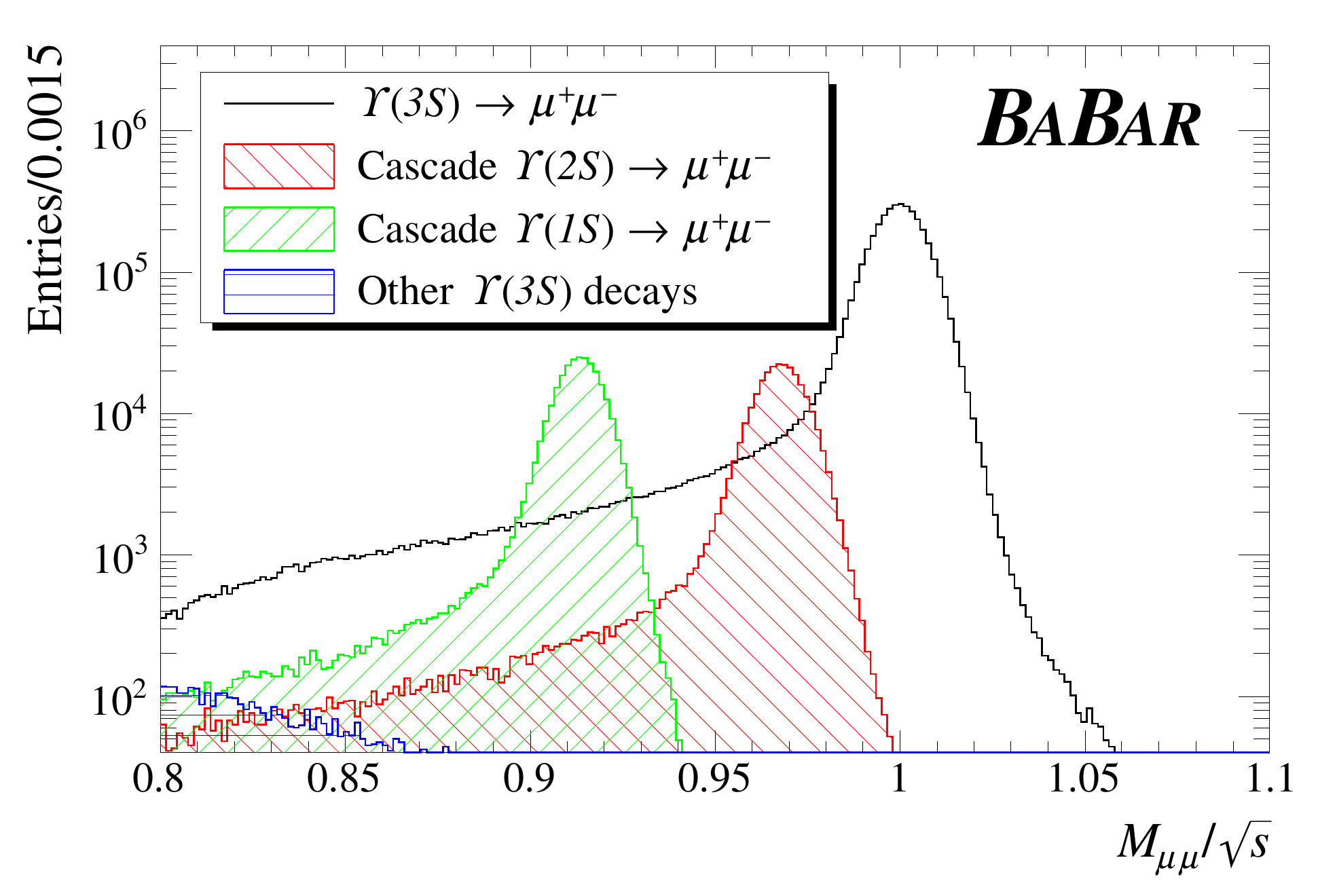}

  \includegraphics[width=0.5\textwidth]{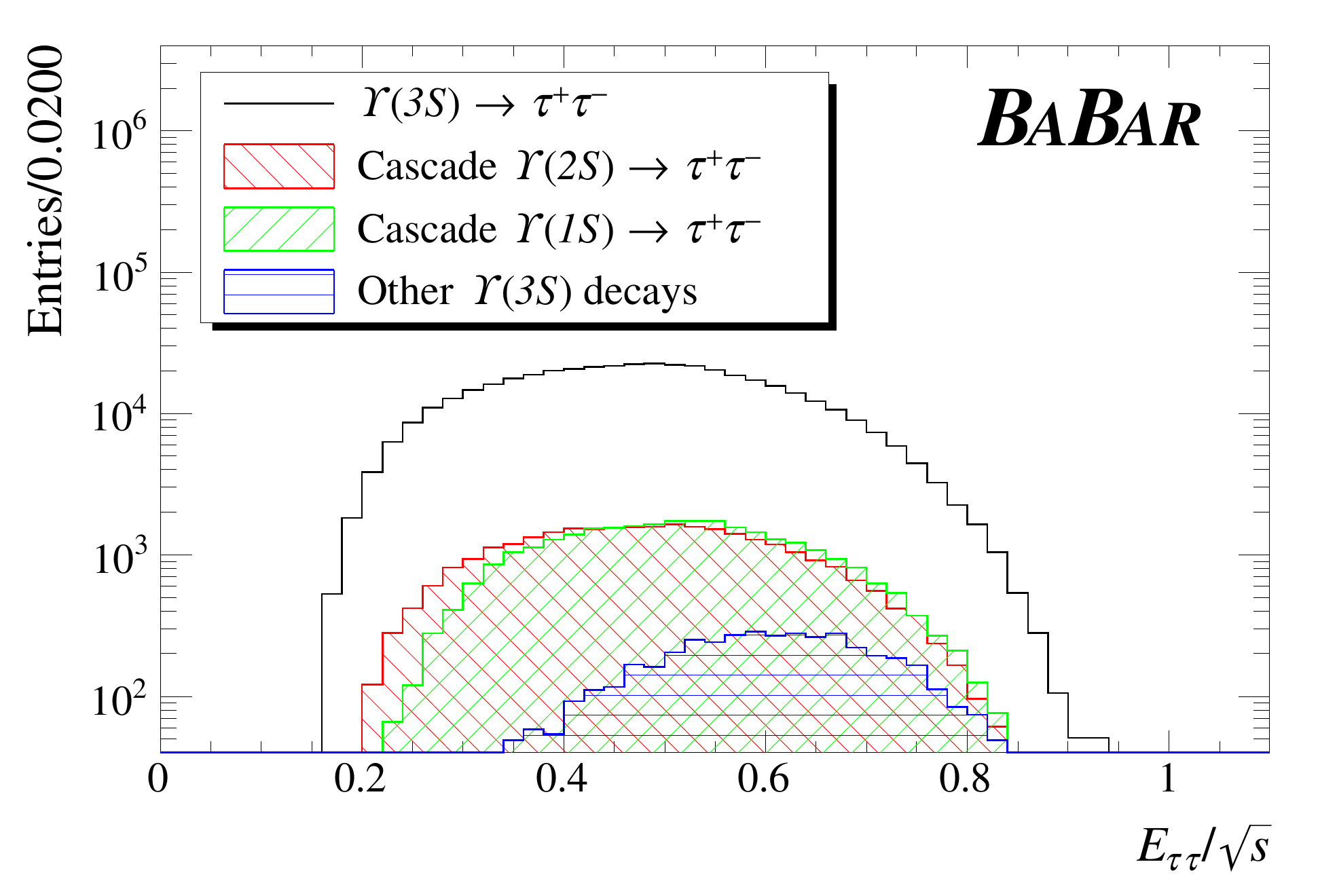}
  \caption{\label{fig:muontautemp} Distributions of
    $M_{\mu\mu}/\sqrt{s}$ (top plot) and $E_{\tau\tau}/\sqrt{s}$
    (bottom plot) variables in MC. Cascade decays are clearly
    separated in dimuon events and nearly indistinguishable in
    $\tau^+\tau^-$ events.}
\end{figure}

In Fig.~\ref{fig:muontautemp} the selected signal events are shown for
simulated $\Upsilon(3S)$ decays. For the dimuon events, the
$M_{\mu\mu}/\sqrt{s}$ variable is plotted whereas for the
$\tau^+\tau^-$ events the total reconstructed event energy, scaled to
center-of-mass energy, $E_{\tau\tau}/\sqrt{s}$, is plotted.  In the
dimuon events, decays of the $\Upsilon(3S)$ to lower mass
$\Upsilon(1S)$ or $\Upsilon(2S)$ resonances via radiative and hadronic
transitions, where the $\Upsilon(1S)$ or $\Upsilon(2S)$ then decay
into a dimuon pair, are clearly seen and separated.  In this paper we
refer to such processes, including analogous processes with a
$\tau^+\tau^-$ final state, as ``cascade decays''.  Owing to the
excellent momentum resolution of the tracking system, the
$M_{\mu\mu}/\sqrt{s}$ distribution provides not only an estimate of
the number of $\Upsilon(3S)\to\mu^+\mu^-$ events but also a direct
evaluation of the contributions from the cascade decays.  In the
$\tau^+\tau^-$ channel, however, these cascade decay channels all have
the same broad distribution in $E_{\tau\tau}/\sqrt{s}$ and are nearly
indistinguishable.

In order to extract the ratio \Rtm that takes into account
correlations between components, a binned maximum likelihood fit
procedure based on the $M_{\mu\mu}/\sqrt{s}$ and
$E_{\tau\tau}/\sqrt{s}$ distributions is employed.  The fit procedure
is based on the template-fitting method of Ref.~\cite{Barlow} and
takes into account the finite number of events in the fit-template
components.  The $\Upsilon(3S)\to\mu^+\mu^-$ and
$\Upsilon(3S)\to\tau^+\tau^-$ templates are taken from the \kkmc-based
MC without ISR effects. The templates for
\mbox{$\Upsilon(2S)\to\ell^+\ell^-$} and
\mbox{$\Upsilon(1S)\to\ell^+\ell^-$} via cascade decays, as well as
the remaining small contributions from $\Upsilon(nS)$ hadronic decays,
are taken from the \evtgen-based MC.  The continuum templates use data
control samples, as described in the following paragraph.

The amount of \babar\ data collected on-resonance is about ten times
larger than off-resonance.  Consequently, when the continuum template
is based only on the off-resonance data, the small size of that sample
dominates the statistical uncertainty of the ratio.  To overcome this
limitation, $\Upsilon(4S)$ on-resonance Run~6 data, with an integrated
luminosity of \SI{78.3}{\per\femto\barn} and the same detector
configuration as Run~7, is used for the continuum template in the fit.
Since the leptonic width of the $\Upsilon(4S)$ is negligible compared
to the total width, only continuum-produced dilepton events are
expected in the sample. However, other $\Upsilon(nS)\to\ell^+\ell^-$
decays appear in the data continuum template via the ISR process.  The
radiative return processes have been extensively studied by
\babar\ (e.g., a narrow resonance production described in
Ref.~\cite{Aubert:2003sv}) and based on this approach, the amount of
ISR-produced $\Upsilon(nS)$ mesons are estimated and subtracted from
the continuum template.

The number of $\Upsilon(3S)\to\mu^+\mu^-$ events $N_{\mu\mu}$ and the
raw ratio $\tilde{R}_{\tau\mu}=N_{\tau\tau}/N_{\mu\mu}$ are free
parameters of the fit. In the non-signal templates, this ratio is
fixed either as in data for the continuum background or to the
simulation prediction for the other templates.

\begin{figure}[htbp!]
  \includegraphics[width=0.5\textwidth]{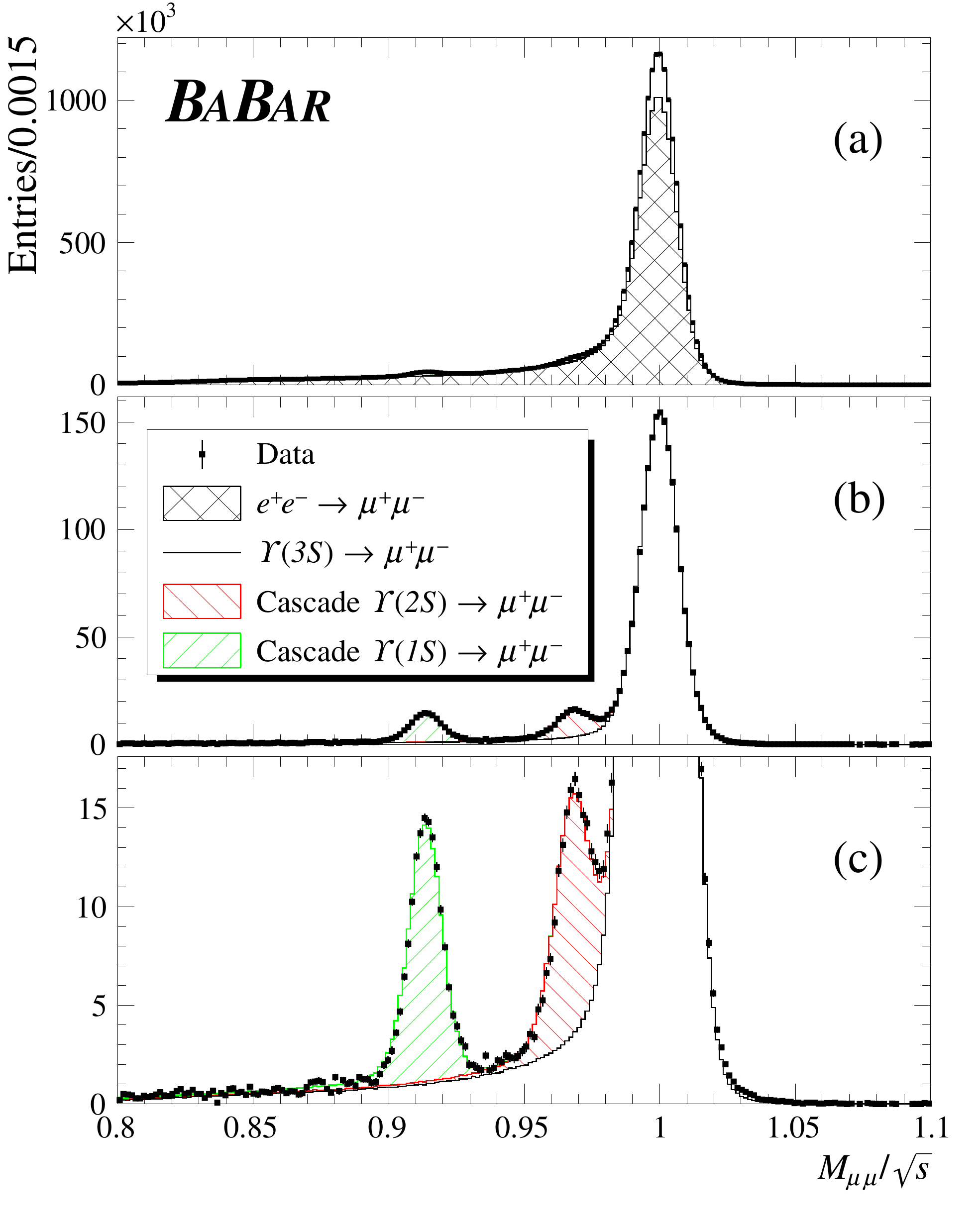}

  \caption{\label{fig:fitmm} The result of the template fit to the
    $\Upsilon(3S)$ data in the $M_{\mu\mu}/\sqrt{s}$ variable.  In (a)
    all events are shown, in (b) and (c) the dominant continuum
    $e^+e^-\to\mu^+\mu^-$ background is subtracted, and (c) is a
    magnified view of (b) to better show cascade decays and the
    radiative tail region.}
\end{figure}

\begin{figure}[htbp!]
  \includegraphics[width=0.5\textwidth]{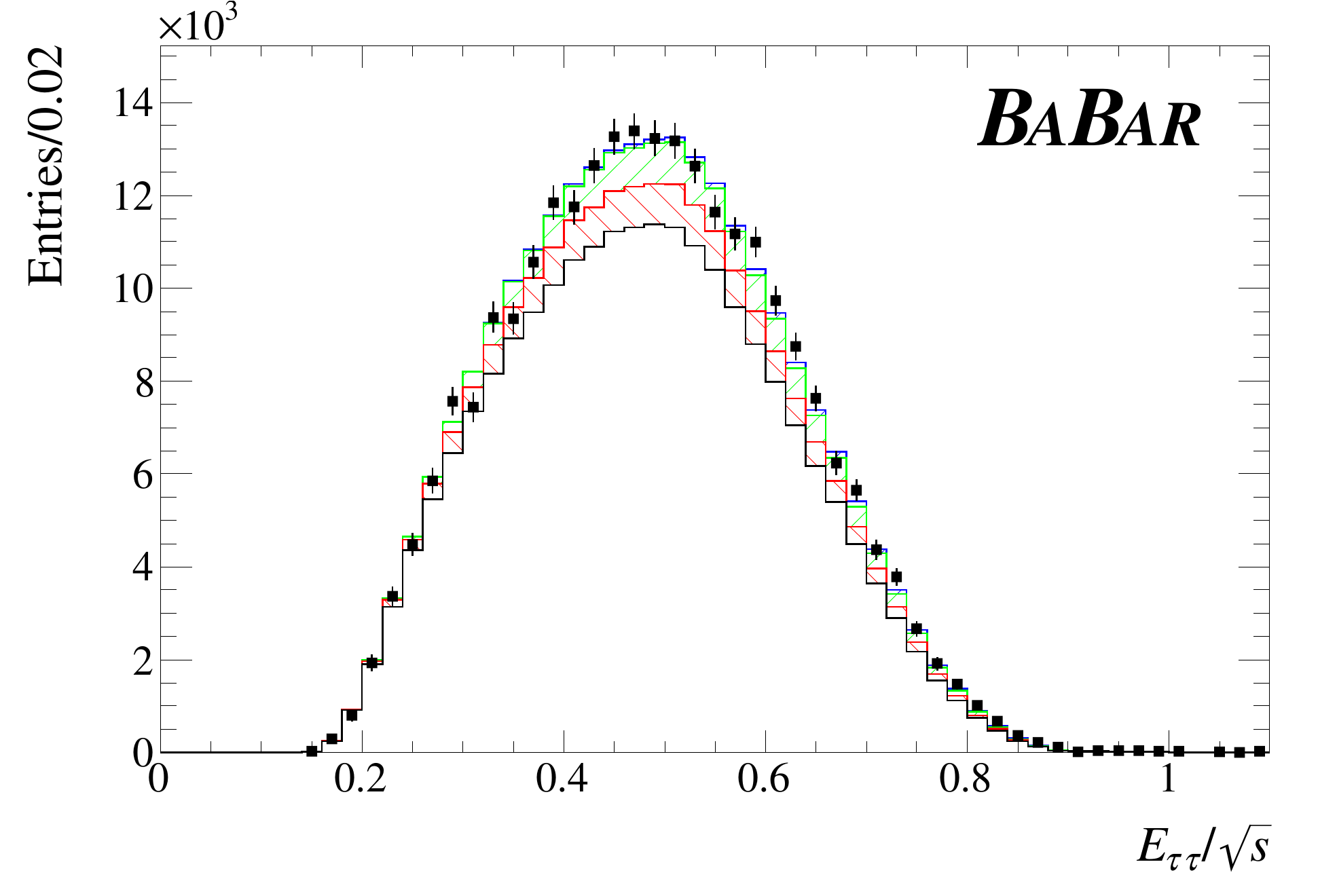}

  \caption{\label{fig:fittt} The result of the template fit to the
    $\Upsilon(3S)$ data in the $E_{\tau\tau}/\sqrt{s}$ variable after the
    continuum background is subtracted.
    Data are depicted  as points with error bars.
    The legend is the same as in
    the corresponding plot in Fig.~\ref{fig:muontautemp}. }
\end{figure}

A graphical representation of the fit result is shown in
Figs.~\ref{fig:fitmm} and \ref{fig:fittt}.  The fit yields a raw ratio
of $\tilde{R}_{\tau\mu}=N_{\tau\tau}/N_{\mu\mu} = 0.10788 \pm
0.00091.$ The MC-based selection efficiencies and their ratio, which
are needed to obtain the ratio $R_{\tau\mu}$, are shown in
Table~\ref{table:eff}.
\begin{table}[htbp!]
  \caption{\label{table:eff} MC selection efficiencies in percent for
    $\Upsilon(3S)\to\ell^+\ell^-$. The quoted uncertainties reflect MC
    statistics.}  \centering
  \begin{tabular}{c|c|c}
    \hline\hline
    $\varepsilon_{\mu\mu}$ (\%) &  $\varepsilon_{\tau\tau}$ (\%) & $\varepsilon_{\tau\tau}$/$\varepsilon_{\mu\mu}$ \\
    \hline
    $69.951 \pm 0.018$ & $7.723 \pm 0.010$ & $0.11041 \pm 0.00015$ \\
    \hline\hline
  \end{tabular}
\end{table}

Low multiplicity $\Upsilon(4S)\to B\bar{B}$ decays, such as
semileptonic decays, can potentially mimic $\tau$-pair events and then
pass the selection criteria. These would modify the
$\Upsilon(4S)$-based continuum template. Note that significant numbers
of $\Upsilon(4S)\to B\bar{B}$ events are not expected in the final
dimuon sample since $M_{\mu\mu}$ of such candidates is too small. To
estimate this effect, a MC sample of $265$ million $\Upsilon(4S)\to
B\bar{B}$ events was processed, which is about three times the size of
the $\Upsilon(4S)$ data, and resulted in 15 dimuon and 7644
$\tau^+\tau^-$ candidates. Thus, the $B\bar{B}$ contribution to the
muon template can be safely neglected whereas the amount of
$\tau^+\tau^-$ candidates translates into a correction of
$\delta_{B\bar{B}}=0.42\%$ to the expected number of
$\Upsilon(3S)\to\tau^+\tau^-$ candidates and is applied to the ratio
$R_{\tau\mu}$.

Combining the fit result $\tilde{R}_{\tau\mu}$, the ratio of MC
efficiencies $\varepsilon_{\mu\mu}/\varepsilon_{\tau\tau}$, the
data/MC correction $C_\text{MC}$, and the correction from $B\bar{B}$
events $\delta_{B\bar{B}}$, the ratio is
$$ \Rtm = \tilde{R}_{\tau\mu} \frac{1}{C_\text{MC}}
\frac{\varepsilon_{\mu\mu}}{\varepsilon_{\tau\tau}}\cdot
(1+\delta_{B\bar{B}}) = 0.9662 \pm 0.0084,$$ where uncertainties from the
data/MC correction and MC efficiencies are included in the statistical
uncertainty.

The sources of the systematic uncertainty in $\Rtm$ are summarized in
Table~\ref{table:syst}.
\begin{table}[b!]
  \caption{\label{table:syst}The summary of systematic
    uncertainties.}  \centering
  \begin{tabular}{c|c}
    \hline\hline
    Source & Uncertainty (\%) \\
    \hline
    Particle identification & 0.9 \\
    Cascade decays & 0.6 \\
    Two-photon production & 0.5 \\
    $\Upsilon(3S)\to\text{hadrons}$ & 0.4\\
    MC shape & 0.4 \\
    $B\bar{B}$ contribution & 0.2 \\
    ISR subtraction & 0.2  \\
    \hline
    Total & 1.4 \\
    \hline\hline
  \end{tabular}
\end{table}

To assess the particle identification uncertainty, three additional
$\tau^+\tau^-$ classifiers were considered. The first used tighter
electron selectors for both the $\tau$ to electron and the $\tau$ to
non-electron selection.  The second had a tighter electron selector
for the $\tau$ to non-electron selection.  The third replaced the
$\tau$ to non-electron selection with an explicit requirement that the
non-electron particle be identified as a muon or a pion.  Even though
the data-driven corrections associated with each of these separate
$\tau^+\tau^-$ classifiers were applied, and despite the highly
correlated statistics in these samples, there remains a 0.9\%
difference between one of these three test classifiers and the default
classifier, which we assign as the particle identification systematic
uncertainty.

The ratio of the number of dimuon and $\tau^+\tau^-$ events from the
cascade decays in the MC fit templates are fixed according to
lepton-flavor universality. To assess the effect of this assumption,
the ratio was varied according to the current experimental
uncertainties in branching fractions for $\Upsilon(1S)$ and
$\Upsilon(2S)$ to dimuon and $\tau^+\tau^-$ final states, resulting in
a maximum difference in $\tilde{R}_{\tau\mu}$ to be 0.6\%, which is
taken as the systematic uncertainty.

As there is no reliable two-photon fusion MC, the contribution to the
systematic uncertainty arising from the two-photon fusion background
is estimated by varying the selection on the transverse momentum,
which reduces the $\tau^+\tau^-$ selection efficiency to almost half
its nominal value.  These variations result in a maximal deviation in
$\tilde{R}_{\tau\mu}$ of 0.5\%.

The simulation of other generic $\Upsilon(3S)$ decays shows that a
small fraction of background events (about 0.1\% of dimuon and 1\% of
$\tau^+\tau^-$ samples) still pass the selection criteria.  These
backgrounds do not exhibit any features that allow them to be easily
separated in the fit itself. Because of this, the amount of this
background is fixed to the MC prediction and a 0.4\% systematic
uncertainty is assessed by varying by 50\% the background, which is
dominated by the $\Upsilon(3S)\rightarrow$hadrons that primarily
contaminate the $\tau^+\tau^-$ sample.

To estimate the systematic uncertainty associated with imperfect
modeling of radiative effects, the \kkmc-based templates for
$\Upsilon(3S)\to\ell^+\ell^-$ decays used in the fit are replaced with
templates created using \evtgen with \photos. This primarily modifies
the shape associated with radiative tail, shown in
Fig.~\ref{fig:fitmm}, resulting in a change in $\tilde{R}_{\tau\mu}$
of 0.2\%.  There is a small $\sim 1$\% difference in $M_{\mu\mu}$
resolution between $\Upsilon(3S)$ and $\Upsilon(4S)$ data, as well as
the same order of magnitude difference between data and MC. To
estimate the systematic uncertainty due to this difference, the mass
resolution in the MC is degraded to be up to 10\% worse than the
resolution in data. This results in a shift in $\tilde{R}_{\tau\mu}$
of up to 0.4\%. From this study, the uncertainty from the MC template
shape mismodeling of $\Upsilon(3S)\to\mu^+\mu^-$ is conservatively
estimated to be 0.4\%. The total systematic uncertainty from the MC
shape modelling associated with the radiative and resolution effects
is 0.4\%.

The uncertainty from the $B\bar{B}$ background in the continuum
template is estimated by varying the expected amount of the background
by 50\%, resulting in a 0.2\% change in the ratio.

The systematic uncertainty associated with $\Upsilon(nS)$ mesons
produced by the radiative return process in the continuum template is
estimated by accounting for experimental uncertainties of total widths
and leptonic branching fractions of these mesons and by varying the
overall amount of these produced mesons by 10\% in order to
conservatively account for radiator function uncertainties.  We assign
a value of 0.2\% as the systematic uncertainty coming from these
various effects.

All of the systematic uncertainties described in the paragraphs above
are combined in quadrature, giving total systematic uncertainty of
1.4\%.

In conclusion, based on the data collected by the \babar\ detector
near the $\Upsilon(3S)$ and $\Upsilon(4S)$ resonances, the ratio of
the leptonic branching fractions of the $\Upsilon(3S)$ meson is
measured to be $$\Rtm = 0.966 \pm 0.008_\text{stat} \pm
0.014_\text{syst}.$$ This is in agreement with the SM prediction of
0.9948~\cite{Aloni:2017eny} within two standard deviations and its
uncertainty almost an order of magnitude smaller than the only
previous measurement reported by the CLEO
collaboration~\cite{Besson:2006gj}.

\label{sec:Acknowledgments}
\input acknow_PRL.tex

\label{sec:bibliography}

\end{document}

%% file: authors_feb2020_frozen.tex
\author{J.~P.~Lees}
\author{V.~Poireau}
\author{V.~Tisserand}
\affiliation{Laboratoire d'Annecy-le-Vieux de Physique des Particules (LAPP), Universit\'e de Savoie, CNRS/IN2P3,  F-74941 Annecy-Le-Vieux, France}
\author{E.~Grauges}
\affiliation{Universitat de Barcelona, Facultat de Fisica, Departament ECM, E-08028 Barcelona, Spain }
\author{A.~Palano}
\affiliation{INFN Sezione di Bari and Dipartimento di Fisica, Universit\`a di Bari, I-70126 Bari, Italy }
\author{G.~Eigen}
\affiliation{University of Bergen, Institute of Physics, N-5007 Bergen, Norway }
\author{D.~N.~Brown}
\author{Yu.~G.~Kolomensky}
\affiliation{Lawrence Berkeley National Laboratory and University of California, Berkeley, California 94720, USA }
\author{M.~Fritsch}
\author{H.~Koch}
\author{T.~Schroeder}
\affiliation{Ruhr Universit\"at Bochum, Institut f\"ur Experimentalphysik 1, D-44780 Bochum, Germany }
% ...
\collaboration{The \babar\ Collaboration}
\noaffiliation

%% file: acknow_PRL.tex
We are grateful for the excellent luminosity and machine conditions
provided by our \pep2\ colleagues, 
and for the substantial dedicated effort from
the computing organizations that support \babar.
The collaborating institutions wish to thank 
SLAC for its support and kind hospitality. 
This work is supported by
DOE
and NSF (USA),
NSERC (Canada),
CEA and
CNRS-IN2P3
(France),
BMBF and DFG
(Germany),
INFN (Italy),
FOM (The Netherlands),
NFR (Norway),
MES (Russia),
MINECO (Spain),
STFC (United Kingdom),
BSF (USA-Israel). 
Individuals have received support from the
Marie Curie EIF (European Union)
and the A.~P.~Sloan Foundation (USA).


%% file: RtaumuLetter.bbl
\begin{thebibliography}{99}
 
\bibitem{VanRoyen:1967nq} 
  R.~Van Royen and V.~F.~Weisskopf,
  Nuovo Cim.\ A {\bf 50}, 617 (1967);
  Erratum: Nuovo Cim.\ A {\bf 51}, 583 (1967).

\bibitem{SanchisLozano:2003ha} 
  M.~A.~Sanchis-Lozano, \ijmpa{19}, 2183 (2004).

\bibitem{Besson:2006gj}
  D.~Besson {\em et al.} (CLEO Collaboration), \prl{\bf 98}, 052002 (2007).

\bibitem{Aloni:2017eny}
  D.~Aloni, A.~Efrati, Y.~Grossman and Y.~Nir, J.\ High\ Energ.\ Phys.\ \textbf{06}, 019 (2017).

\bibitem{Amhis:2016xyh}
  Y.~Amhis \textit{et al.} (HFLAV Group), \epjc{77}, no.12, 895 (2017).

\bibitem{lumi}
  J.P. Lees {\em et al.} ({\babar} Collaboration), \nima{726}, 203 (2013).
  
\bibitem{ref:blindanalysis}
  J.~R.~Klein and A.~Roodman, \arnps{55}, 141 (2005).

\bibitem{ref:aubert}
 B. Aubert {\em et al.}  (\babar\ Collaboration), \nima{479}, 1 (2002).

\bibitem{ref:NIMUpdate}
B. Aubert {\em et al.}  (\babar\ Collaboration), \nima{729}, 615 (2013).

\bibitem{ref:ward}
B.~F.~L.~Ward, S.~Jadach and Z.~Was, \npps{116}, 73 (2003).

\bibitem{ref:BHWIDE}
S.~Jadach, W.~Placzek,  B.~F.~L.~Ward, \plb{390}, 298 (1997).

\bibitem{ref:lange}
  D.~J.~Lange, \nima{462}, 152 (2001).

\bibitem{photos} 
  E.~Barberio and Z.~Was, \cpc{79}, 291 (1994).

\bibitem{ref:agostine}
  S. Agostinelli {\em et al.} (\geant4 Collaboration), \nima{506}, 250 (2003).

\bibitem{Barlow}
  R.~J.~Barlow and C.~Beeston, \cpc{77}, 219 (1993).

\bibitem{Aubert:2003sv} 
  B.~Aubert {\em et al.}  (\babar\ Collaboration), \prd{\bf 69}, 011103 (2004).
  
 \end{thebibliography}
